\documentclass[aps,showpacs,prb,twocolumn,floatfix,superscriptaddress]{revtex4}
\usepackage[english]{babel}
\usepackage{graphicx}
\usepackage{color}
\usepackage[utf8]{inputenc}
\usepackage{pstricks,pst-grad,color}
\usepackage{graphicx,amssymb}
\usepackage{amsmath}
\usepackage{amssymb}
\usepackage{bbold}
\usepackage{placeins}
\usepackage{bm}
\usepackage{comment}

\begin{document}

\title{Quantum oscillations in strongly correlated topological Kondo insulators}

\begin{abstract}
The observation of quantum oscillations in topological Kondo insulators SmB$_6$ and YbB$_{12}$ is a recent puzzling experimental discovery. 
Quantum oscillations observed in the resistivity and the magnetization are usually explained by the existence of the Fermi surface. However, Kondo insulators do not have a Fermi surface and thus should not show quantum oscillations.
By performing dynamical mean field  calculations for  topologically nontrivial Kondo insulators in a magnetic field, we analyze the effect of correlations on the emergence of quantum oscillations in narrow-gap topological Kondo insulators and demonstrate that the interplay between correlations and nonlocal hybridization, ubiquitously occurring in topological Kondo insulators, can lead to observable quantum oscillations without the necessity of a Fermi surface.
Particularly, we show that correlations make it easier to observe quantum oscillations  in the magnetization and the resistivity of the bulk material. 
The fundamental mechanism for these quantum oscillations 
is a combination of correlation effects and Landau levels coming very close to the Fermi energy. 
We furthermore demonstrate that quantum oscillations in a three-dimensional system can be understood by analyzing the physics on the two-dimensional planes in the momentum space for which the hybridization in direction of the magnetic field vanishes.
We believe that this scenario  is relevant to understanding the observation of quantum oscillations in the magnetic torque for SmB$_6$ as well as oscillations in the resistivity and the magnetic torque of YbB$_{12}$.
\end{abstract}

\author{Robert Peters}
\email[]{peters@scphys.kyoto-u.ac.jp}
\affiliation{Department of Physics, Kyoto University, Kyoto 606-8502, Japan}

\author{Tsuneya Yoshida}
\affiliation{Department of Physics, University of Tsukuba, Ibaraki 305-8571, Japan }

\author{Norio Kawakami}
\affiliation{Department of Physics, Kyoto University, Kyoto 606-8502, Japan}

\newcommand*{\tran}{^{\mkern-1.5mu\mathsf{T}}}
\newcommand{\1}{\mbox{1}\hspace{-0.25em}\mbox{l}}
\date{\today}



\maketitle
\section{introduction}
One of the most puzzling recent experimental discoveries in condensed matter physics is the observation of quantum oscillations in insulating materials SmB$_6$ and YbB$_{12}$\cite{Tan287,Xiangeaap9607,Ong32,0953-8984-30-16-16LT01,Hartstein2017}. 
While firstly these oscillations have been attributed to metallic surface states\cite{Li1208}, there is also strong evidence that they arise from the bulk\cite{Tan287,Xiangeaap9607,Ong32,0953-8984-30-16-16LT01,Hartstein2017,PhysRevLett.116.046403}.
Both materials are strongly correlated $f$ electron systems for which a gap develops because of a hybridization between conduction ($c$) electrons and strongly correlated $f$ electrons, and thus a large resistivity at low temperatures can be measured \cite{Fisk1995798,Riseborough2000}. Surprisingly, quantum oscillations in strong magnetic fields have been observed in SmB$_6$ in the magnetic torque\cite{Tan287} and in YbB$_{12}$ in the resistivity and the magnetic torque\cite{Xiangeaap9607}. These measurements contradict our  understanding of quantum oscillations, which
is rooted in the existence of a Fermi surface; electron bands, which create the Fermi surface, form Landau levels in a magnetic field. When the magnetic field strength is changed, the energies of these Landau levels change, leading to an oscillatory behavior in most of the observable quantities. However, insulating materials such as SmB$_6$ and YbB$_{12}$ do not possess a Fermi surface, thus there are no electrons, which can form Landau levels, close to the Fermi energy.
Because of this discovery, different theories have been used to explain these observations: quantum oscillations might be observed under special conditions, if the gap is very small\cite{PhysRevLett.115.146401,PhysRevB.94.125140,PhysRevB.95.085111,PhysRevB.96.235121,PhysRevB.99.045149,PhysRevB.96.075115,PhysRevB.96.195122}. Other theories explain these quantum oscillations by composite-excitons\cite{PhysRevLett.118.096604,Chowdhury2018} or Majorana-fermions\cite{PhysRevLett.119.057603}, which form a Fermi surface. However, the existence of fermionic charge-neutral excitations is highly controversial.\cite{PhysRevLett.116.246403}
Furthermore, it has been pointed out that quantum oscillations can be understood using non-hermitian properties of the material\cite{PhysRevLett.121.026403}.
Besides these works, in a study for narrow-gap topological insulators, Zhang {\it et al.}\cite{PhysRevLett.116.046404} have shown in a noninteracting continuum model that the gap of a topological insulator closes at a critical magnetic field, which has been further analyzed in other works.\cite{PhysRevB.97.115202} They have shown that this gap closing is accompanied by oscillatory behavior in observable quantities.
In this way, there are a variety of  theories based on different assumptions, which might explain these quantum oscillations.
However, a conclusive answer has not been found yet, although this problem is of uttermost importance: Quantum oscillations are viewed as an accurate method for the experimental determination of the Fermi surface of materials, and the measurement of quantum oscillations in an insulator contradict the existing theories.

The only examples of insulating material showing quantum oscillations in magnetic fields are SmB$_6$ and YbB$_{12}$, which are both good candidates for topological Kondo insulators.\cite{Takimoto2011,PhysRevLett.110.096401,PhysRevLett.111.226403,PhysRevB.85.045130,PhysRevLett.104.106408,PhysRevLett.112.016403,annurev-conmatphys-031214-014749,PhysRevB.93.235159}
Thus, it is natural to ask, whether nontrivial topology, strong correlations, or a combination of both is essential to observe quantum oscillations in insulating materials.

We here answer this question by analyzing correlation effects on quantum oscillations appearing in the bulk of a two-dimensional (2D) topological Kondo insulator. 
(i) We confirm in our work that the gap closing due to a momentum-dependent hybridization as described by  Zhang {\it et al.}\cite{PhysRevLett.116.046404} for a noninteracting continuum model exists in strongly correlated lattice models.  Such a momentum-dependent hybridization is ubiquitously found in topological Kondo insulators.
(ii) We show that not only the gap-width is reduced by the renormalization of the band structure, but also the slope of the gap-closing with increasing magnetic field strength decreases. Thus, the critical field strength, where the gap closes and the insulator changes into a metal, is nearly unchanged. 
(iii) We demonstrate that the amplitude of the quantum oscillations is enhanced by correlations due to the finite life-time of quasi-particles.  We believe that this together with the renormalization of the band structure is an important fact for the experimental detection and might explain why quantum oscillations in insulating materials have been so far only observed in strongly interacting topological materials.
(iv) Finally, we demonstrate that the quantum oscillations occurring in the bulk of a three-dimensional (3D) system are determined by the momenta where the hybridization in the direction of the magnetic field vanishes, and that the physics at the Fermi energy of the 3D bulk is essentially the same as in 2D.
This validates the relevance of our study to 3D materials such as  SmB$_6$ and YbB$_{12}$.

The rest of this paper is organized as follows: In the next section, Sec. \ref{sec_model}, we introduce the model and methods used to analyze the quantum oscillations in a topological Kondo insulator. This is followed by a section about the gap closing in the noninteracting model, Sec. \ref{sec_noninter}. The main results are shown in  Sec. \ref{sec_inter},  where we analyze quantum oscillations due to the interplay of strong correlations and nontrivial topology.  
In Sec. \ref{threeD}, we show results for a 3D model and explain the relation and relevance of the 2D model to 3D systems.
Finally, in Sec. \ref{sec_dis}, we discuss the obtained results and the relevance for SmB$_6$ as well as for YbB$_{12}$ and conclude the paper.

\section{Model and method\label{sec_model}}
\subsection{Model}
Because both materials, SmB$_6$ and YbB$_{12}$, have different band structures, we believe that 
the details of the band structure may not play a major role for explaining these quantum oscillations. 
To obtain generic properties, we thus study a model of a two-dimensional topological Kondo insulator which captures the essence of the interplay between correlations and nontrivial topology. 

Our model includes one $f$-electron- and one $c$-electron-band, which hybridize via a nonlocal hybridization.
This results in a topologically nontrivial gap at the Fermi energy.
We furthermore include a strong local Coulomb interaction in the $f$-electron band. The model is thus a periodic Anderson model with a momentum dependent hybridization,
and can be written as
\begin{eqnarray}
H&=&\sum_{\vec k,\sigma}\left(\epsilon^c_{\vec k}c^\dagger_{\vec k,\sigma} c_{\vec k, \sigma}+\epsilon^f_{\vec k}f^\dagger_{\vec k,\sigma} f_{\vec k, \sigma}\right)+\sum_i Uf^\dagger_{i,\uparrow}f_{i,\uparrow}f^\dagger_{i,\downarrow}f_{i,\downarrow}+\nonumber\\
&&\sum_{i,\sigma} \left(\mu_f f^\dagger_{i,\sigma}f_{i,\sigma}+\mu_c c^\dagger_{i,\sigma}c_{i,\sigma}\right)+H_{\text{Hyb}}\label{Ham},
\end{eqnarray}
with
\begin{eqnarray}
H_{\text{Hyb}}&=&\sum_{\vec k}2V\left((\sigma_x)_{\rho_1\rho_2} \sin k_x c^\dagger_{\vec k,\rho_1}f_{\vec k,\rho_2} +\text{h.c.}\right)+\nonumber\\
&&\sum_{\vec k}2V\left((\sigma_y)_{\rho_1\rho_2} \sin k_y c^\dagger_{\vec k,\rho_1}f_{\vec k,\rho_2} +\text{h.c.}\right)\label{eq_hyb}.
\end{eqnarray}
\begin{figure}[t]
\begin{center}
    \includegraphics[width=\linewidth]{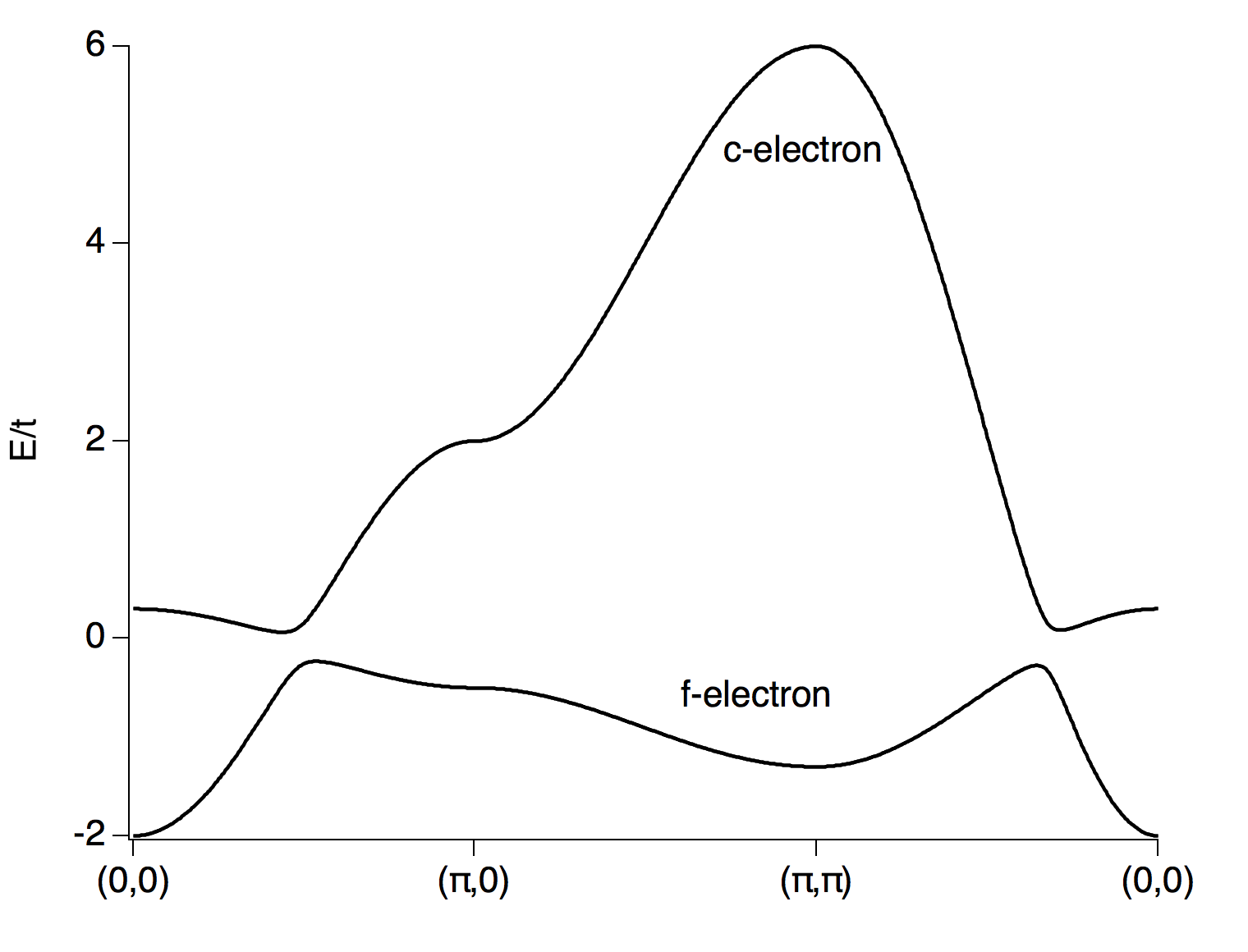}  
\end{center}
\caption{Noninteracting band structure showing the $f$- and $c$-electron bands and the resulting gap at the Fermi energy, $E/t=0$. The chemical potentials $\mu_c$ and $\mu_f$ are chosen in a way that $n_f=1.5$, thus $n_c=0.5$ in the valence fluctuating regime.
\label{Fig1}}
\end{figure}
The band structure corresponds to  a 2D tight binding model with only nearest neighbor hopping. The dispersion reads $\epsilon^{c/f}_{\vec k}=2t_{c/f}(\cos k_x +\cos k_y)$. We choose the hopping of the $c$ electrons, $t_c=t$, as unit of energy throughout this paper and set the hybridization strength to $V/t=0.1$ and $t_f/t=0.2$.
$H_{\text{Hyb}}$ describes the hybridization between $f$ and $c$ electrons, which results in a topologically nontrivial gap at the Fermi energy. Besides a hybridization arising from spin-orbit interaction, we need to change the $f$ electron number away from $n_f=1$ in order to obtain a topological insulator.
In the absence of a magnetic field, the analyzed model has only time-reversal symmetry and thus can be a topological insulator of class AII where systems may have a $Z_2$ topological invariant in two and three dimensions.\cite{RevModPhys.88.035005} 
Because our model has furthermore inversion symmetry, the topological invariant can be easily calculated in the noninteracting model by the product of the eigenvalues of the parity operator for the occupied states at the time-reversal invariant momenta in the Brillouin zone.\cite{PhysRevB.76.045302} We find that this product is negative, corresponding to a topologically nontrivial model. 
We note that the $Z_2$ invariant for correlated systems can be defined by the single-particle Green's function as long as it is nonsingular (i.e., there is neither gap-closing nor a divergence of the self-energy).\cite{PhysRevX.2.031008} We have confirmed that the $Z_2$ invariant remains nontrivial even in the strongly correlated region. 
Besides verifying these conditions, we also have directly calculated the surface spectrum without magnetic field and have confirmed the existence of metallic surface states at $\vec k=(0,0)$. We thus confirm that the analyzed model is a topological Kondo insulator. 
In Fig. \ref{Fig1}, we show a typical band structure for the valence fluctuating regime with $n_f=1.5$ and $n_c=0.5$. Clearly visible is a flat $f$ electron band and a wide $c$ electron band. Furthermore, at the Fermi energy $E/t=0$, a gap exists.

\subsection{Peierls phase on a lattice}
\begin{figure}[t]
\begin{center}
    \includegraphics[width=\linewidth]{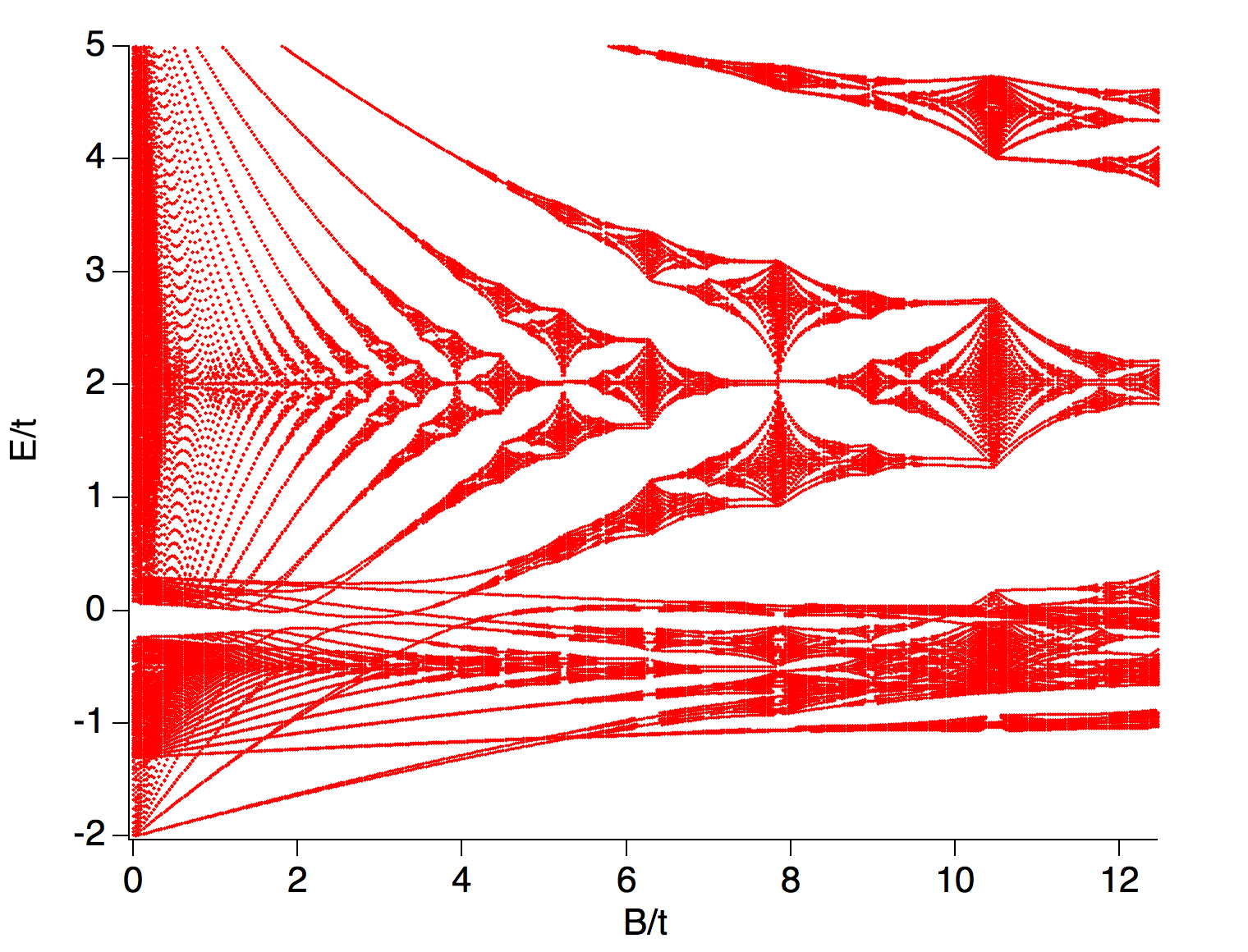}  
\end{center}
\caption{Energy diagram in the noninteracting system for $V/t=0.1$ showing a typical Hofstadter-butterfly structure.
\label{Fig2}}
\end{figure}
Throughout this paper, we consider a magnetic field $\vec B=B(0,0,1)$ in $z$-direction.
When a magnetic field is applied to electrons on  a lattice, the Hamiltonian must be modified using the Peierls substitution\cite{Peierls1933}, which takes into account that electrons moving in a closed path acquire a phase proportional to the magnetic flux through the area enclosed by the path.
We here describe a system with a vector potential $\vec A=B(-y,0,0)$ which leads to a magnetic field in $z$-direction with strength $B$.
To incorporate a vector potential into a tight binding model, we include the appropriate Perierls phases
\begin{subequations}
\begin{eqnarray}
\phi^x_{x,y}&=&\int A_x(x,y)dx,\quad\phi^y_{x,y}=\int A_y(x,y)dy,\\
&&c^\dagger_{x,y}c_{x+1,y}\rightarrow c^\dagger_{x,y}c_{x+1,y}\exp(i\phi^x_{x,y}),\\
&&c^\dagger_{x,y}c_{x,y+1}\rightarrow c^\dagger_{x,y}c_{x,y+1}\exp(i\phi^y_{x,y}),
\end{eqnarray}
\end{subequations}
where $A_x(x,y)$ and $A_y(x,y)$ are given by the vector potential at a lattice site $(x,y)$ in $x$ and $y$ directions, respectively.
Thus, the hopping in $x$ ($y$) direction must be modified in a magnetic field by an additional factor $\exp(i\phi^x_{x,y})$ [$\exp(i\phi^y_{x,y})$].

For a homogeneous magnetic field with vector potential $A=B(-y,0,0)$, we only need to modify the hopping in $x$ direction by a phase factor which depends on the $y$ coordinate of the lattice site, reading
\begin{eqnarray}
H&=&t\sum_{x,y}c^\dagger_{x+1,y}c_{x,y}+c^\dagger_{x,y+1}c_{x,y}+\text{h.c.}\nonumber\\
\Rightarrow H&=&t\sum_{x,y}\left(e^{iBy}c^\dagger_{x+1,y}c_{x,y}+c^\dagger_{x,y+1}c_{x,y}+\text{h.c.}\right)\nonumber,
\end{eqnarray}
where the first (second) equation describes the system without (with) magnetic field.
We thus need to take into account that the wavefunctions on lattice sites with different $y$-index are different. However, if $B=2\pi\frac{M}{N}$ with some integers $M$ and $N$, the wavefunctions on lattice sites with $y$ and $y+N$ are equal, because the phase factors are equal. We thus  make an ansatz for the wavefunction for every lattice site $y=1\ldots N$ 
\begin{displaymath}
\Psi_{x,y}=\exp(ik_x x)\exp(ik_y y)\phi_y\quad\text{with}\quad\phi_{y+N}=\phi_y,
\end{displaymath}
and enlarge the unit cell to the magnetic unit cell including $N$ different lattice sites along the $y$-direction.

Our original Hamiltonian can be written as
\begin{widetext}
\begin{eqnarray}
H&=&\sum_y\sum_{k_x,\sigma}\left(2t_c\cos(k_x-y\phi) c^\dagger_{y,\vec k,\sigma} c_{y,\vec k, \sigma}+2t_f\cos(k_x-y\phi)f^\dagger_{y,\vec k,\sigma} f_{y,\vec k, \sigma}\right)+\nonumber\\
&&\sum_y\sum_{k_y,\sigma}\left(t_c  \exp (ik_y)c^\dagger_{y,\vec k,\sigma} c_{y+1,\vec k, \sigma}+t_f\exp(ik_y)f^\dagger_{y,\vec k,\sigma} f_{y+1,\vec k, \sigma}+\text{h.c.}\right)+\nonumber\\
&&\sum_y\sum_{k_x,\sigma}\left(2V\sin(k_x-y\phi) f^\dagger_{y,\vec k,\uparrow} c_{y,\vec k, \downarrow}+2V\sin(k_x-y\phi) f^\dagger_{y,\vec k,\downarrow} c_{y,\vec k, \uparrow}+\text{h.c.}\right)+\nonumber\\
&&\sum_y\sum_{k_y,\sigma}\left(V  \exp (ik_y)f^\dagger_{y,\vec k,\uparrow} c_{y+1,\vec k, \downarrow}-V\exp(ik_y)f^\dagger_{y,\vec k,\downarrow} c_{y+1,\vec k, \uparrow}+\text{h.c.}\right)+\nonumber\\
&&\sum_y\sum_{k_y,\sigma}\left(V  \exp (ik_y)c^\dagger_{y,\vec k,\uparrow} f_{y+1,\vec k, \downarrow}-V\exp(ik_y)c^\dagger_{y,\vec k,\downarrow} f_{y+1,\vec k, \uparrow}+\text{h.c.}\right),
\end{eqnarray}
\end{widetext}
where $\vec k=(k_x,k_y)$.

This Hamiltonian with periodic boundary conditions enables us to calculate properties for magnetic field strengths which can be written as $B=2\pi\frac{M}{N}$ where $M$ and $N$ are integers. For a tight binding model with a single band on a square lattice, the energy dispersion yields the famous Hofstadter butterfly spectrum. For our model consisting of $4$ different bands (including the spins of the $c$ and $f$ electrons), we obtain an energy diagram which is the overlap of two Hofstadter butterflies, see Fig. \ref{Fig2}.

\subsection{Boundary conditions}
As explained above, for a given magnetic  field $B=2\pi\frac{M}{N}$ ($M$ and $N$ are integers), the unit cell should be enlarged to $N$ lattice sites. As a consequence, only calculations for magnetic fields where the magnetic field is a rational number $\frac{M}{N}$ of $2\pi$ can be exactly simulated. Furthermore, smoothly varying the magnetic field strength is not possible, because the magnetic unit cell changes. Finally, if the magnetic unit cell becomes too large, numerical calculations  become impossible due to limited computer resources. 

We thus choose here a different approach. Instead of simulating a bulk system for which the system size corresponds to the magnetic unit cell, we simulate an open boundary system consisting of $80$ lattice sites. Because the lattice has open boundaries (there are no periodic boundary conditions), the magnetic flux through the whole lattice is not necessarily a multiple of $\frac{2\pi}{80}$, but can be any value. We note here that one can expect deviations from the bulk calculations particularly when the magnetic field strength is very weak, implying very large magnetic unit cells.
 Details and comparisons between  calculations with different boundary conditions are shown in the appendix. There are no qualitative differences for the observation of quantum oscillations between calculations with different boundary conditions.

\subsection{Correlations}

We incorporate correlation effects via 
 the real-space dynamical mean field theory (RDMFT), which is an extension of the dynamical mean field theory\cite{Georges1996} to inhomogeneous situations. The inhomogeneity  arises here from the Peierls phases of the magnetic field 
 leading to a large magnetic unit cell. 
 
  To perform the real-space DMFT calculations, we calculate the local Green's function via
 \begin{equation}
 G(\omega)_{ii}=\int d\vec k\left (\omega+i0-H(\vec k)-\Sigma(\omega)\right)_{ii}^{-1},
 \end{equation}
 where $i$ is the index of the site, $\omega$ the frequency, $H(\vec k)$ the one-particle part of the Hamiltonian of the finite slab or the magnetic unit cell including all Peierls phases, and $\Sigma(\omega)$ a matrix containing the self-energies of all lattice sites. In the first iteration of the real-space DMFT, this self-energy can be set to zero. 
For the system consisting of $80$ layers, we thus find $80$ local Green's functions. For a periodic system with magnetic unit cell of $N$ sites, we have $N$ different Green's functions. 
 From these local Green's functions, we set up independent impurity models by determining the hybridization functions\cite{Bulla2008}
 \begin{equation}
 \Delta_{ii}(\omega)=\omega+i0-G_{ii}^{-1}-\Sigma_{ii}(\omega).
 \end{equation}
 These impurity models are solved using the numerical renormalization group (NRG)\cite{Bulla2008} from which the self-energy for each lattice site is calculated.
 These self-energies are then used to calculate new local Green's functions of all lattice sites. This is iterated until self-consistency is reached.

The usage of NRG enables us to calculate self-energies on the  real-frequency axis with high precision around the Fermi energy\cite{Peters2006}. 
NRG uses a logarithmic discretization of bath states to iteratively diagonalize the impurity Hamiltonian. This logarithmic discretization results in a poor resolution away from the Fermi energy but a high resolution around the Fermi energy. Because the observed quantum oscillations are caused by Landau levels close to the Fermi energy, it is expected that NRG yields accurate results. To confirm that our results are not influenced by the discretization, we have performed test calculations in a magnetic field with different discretization parameters and have found no discernible difference.
All calculations are done for $T/t=10^{-4}$ which is far below the Kondo temperature and the gap width of the model. 
We note that strong correlations can lead to the emergence of long-range ordered magnetic phases in the periodic Anderson model besides the appearance of a ferromagnetic polarization due to the magnetic field. Possible phases are antiferromagnetic phases or charge ordered phases, as has been shown in previous papers\cite{PhysRevB.98.075104}.
Such phase transitions can also be triggered by including the magnetic field, particularly because the gap width is suppressed by the magnetic field. As we are interested here in the paramagnetic phase only showing the polarization due to the magnetic field, we suppress such metamagnetic transitions.

\section{Gap closing mechanism in the noninteracting model\label{sec_noninter}}
\begin{figure}[t]
\begin{center}
  \includegraphics[width=0.9\linewidth]{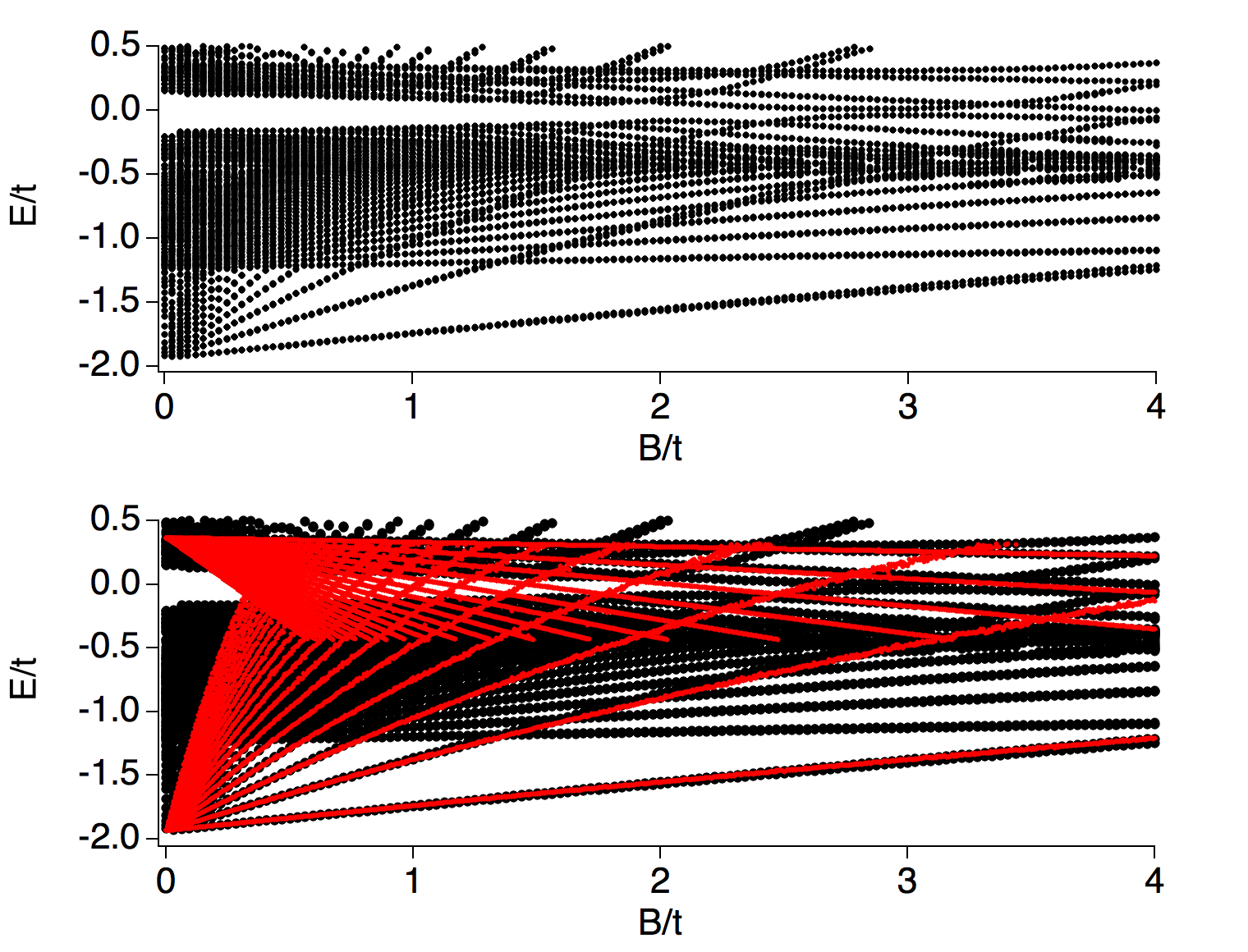}  
\end{center}
\caption{Energy diagram for the noninteracting model for $V/t=0.1$. The bottom panel additionally includes Landau levels described by Eq. (\ref{EQ:LandauLevel}), which agree with the energy diagram of the top panel.
\label{Fig3}}
\end{figure}
\begin{figure}[t]
\begin{center}
  \includegraphics[width=0.9\linewidth]{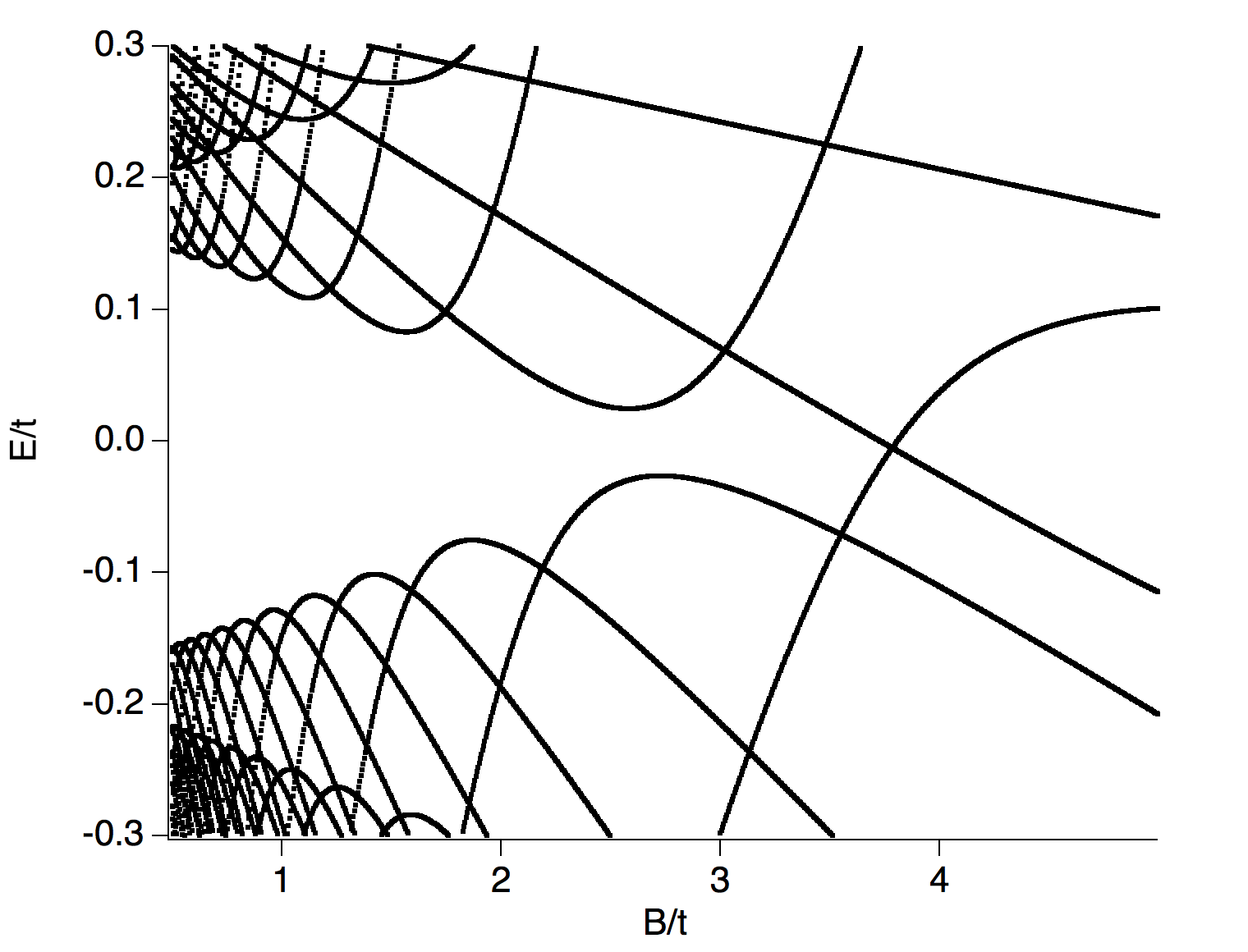}  
\end{center}
\caption{Energy diagram calculated by including a hybridization between Landau levels described by Eq. (\ref{EQ:LandauLevel}).
\label{Fig4}}
\end{figure}
In this section, we firstly analyze the noninteracting model to demonstrate that the gap closes in strong magnetic fields.
The gap closing mechanism in the noninteracting model for strong magnetic fields is equivalent to the situation described in the work of Zhang {\it et al.}\cite{PhysRevLett.116.046404} in a noninteracting continuum model.
In Fig. \ref{Fig3} (top), we show the energy diagram for $n_f=1.5$ in the noninteracting model. In this plot, energy levels approaching the Fermi energy with nearly linear magnetic field dependence are clearly visible. These energy levels can be understood as Landau levels of the $c$ and $f$ electrons.
 We can write these as
 \begin{subequations}
\begin{eqnarray}
E_l^c(B)&=&D_c+\frac{2\pi B}{\partial A^c(E)/\partial E}(l+1/2),\\
E_l^f(B)&=&D_f-\frac{2\pi B}{\partial A^f(E)/\partial E}(l+1/2)\label{EQ:LandauLevel},
\end{eqnarray}
\end{subequations}
where we use $D_c$ and $D_f$ as fitting parameters for $c$ and $f$ electrons, and $A_{c,f}(E)$ is the area of the Fermi surface at energy $E$ for the $c$ and $f$ electrons. Due to the energy dependence of the derivative $\frac{\partial A^{c,f}(E)}{\partial E}$, the dispersion of the Landau levels is bent. In the bottom panel of Fig. \ref{Fig3}, we include these Landau levels into the energy diagram. In this figure, we do not include a hybridization between these $c$ and $f$ electron Landau levels. Thus, these Landau levels do not form a hybridization gap at the Fermi energy. Figure \ref{Fig3} (bottom) shows a good agreement, away from the Fermi energy between the calculated energy diagram of the lattice model and these Landau levels.

To understand the formation and the closing of the hybridization gap, we need to include a hybridization between the $c$ and $f$ electron Landau levels.
In contrast to $f$ electron systems with local hybridization, the hybridization originating in the spin-orbit interaction is an odd function of the momentum such as $\sigma_i\sin(k_i)$, which
leads to a hybridization of Landau levels $l$ with $l+1$ or $l$ with $l-1$.\cite{PhysRevLett.116.046404} The hybridization between Landau levels with different index can explain why the gap closes at strong magnetic fields. In Fig. \ref{Fig4}, we use the Landau levels from Fig. \ref{Fig3} and include a hybridization between Landau levels with different index as described above. 

Assuming that the gap opens when the energetic distance between the $c$ and the $f$ electron Landau levels is of the size of the hybridization, i.e. $V \approx E_l^c(B)-E_{l\pm 1}^f(B)$, we can derive
\begin{eqnarray}
B_l&=&\left \vert \frac{(V-D_c+D_f)}{(\frac{2\pi}{\partial A^c(E)/\partial E}(l+1/2)-\frac{2\pi}{\partial A^f(E)/\partial E}(l+1/2\pm 1))}\right \vert.\nonumber\\
\end{eqnarray}
Solving this equation for $l$ and inserting it into the energy equation of the Landau levels, we obtain the energy depending on the magnetic field where the hybdrization gaps open. The dependence of the energy on the magnetic field reads
\begin{equation}
E\sim\frac{2\pi}{\partial A_c(E)+\partial A_f(E)}B=\frac{1}{m_c+m_f}B.
\end{equation}
Thus, the slope of the gap closing in the energy diagram of the lattice model is proportional to the inverse of the mass of the electrons.

\section{quantum oscillations in the strongly correlated model\label{sec_inter}}
\begin{figure}[t]
\begin{center}
    \includegraphics[width=0.9\linewidth]{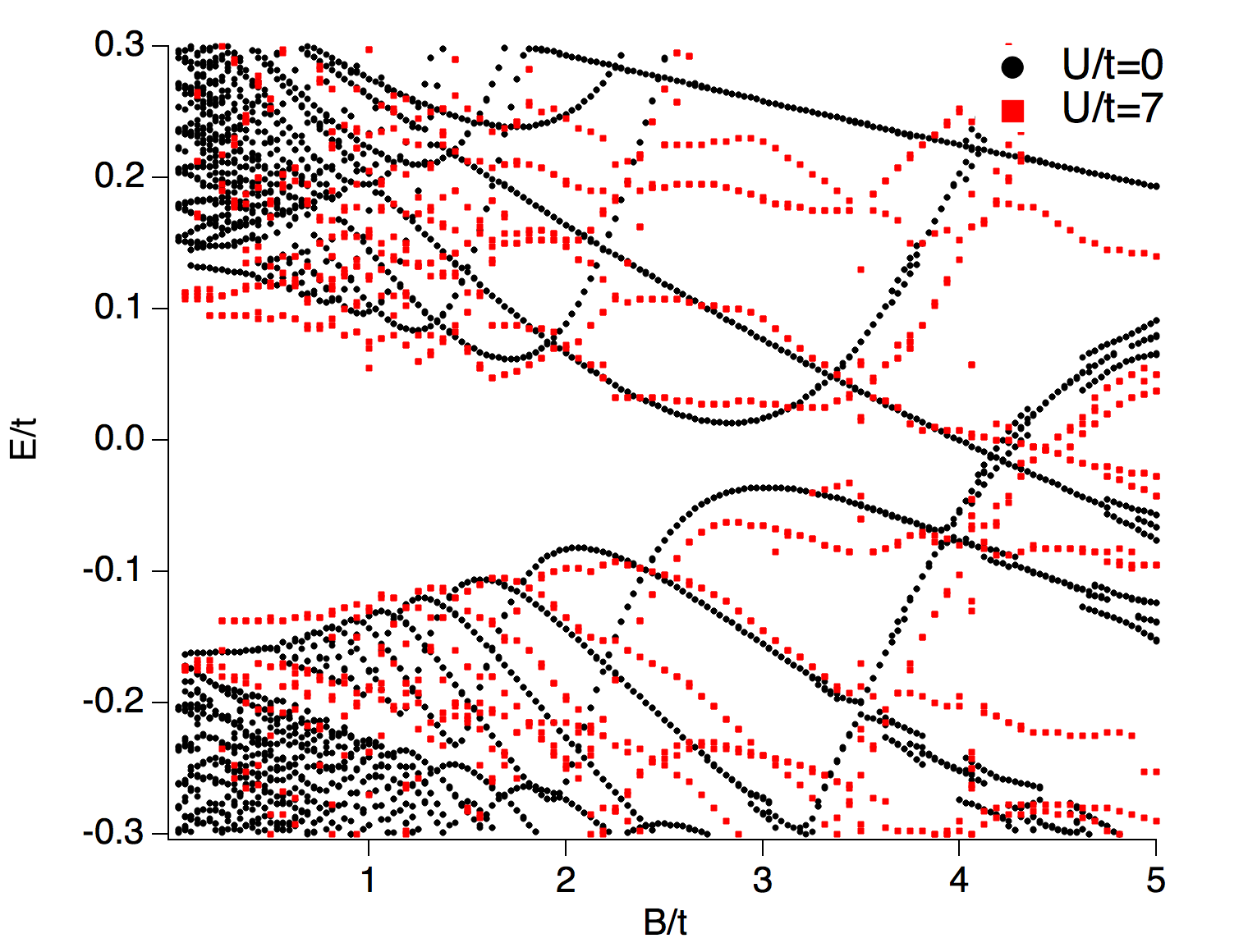}  
\end{center}
\caption{Energy diagram in the valence fluctuating regime, $n_f=1.5$. Comparison of the energy level structure between the noninteracting and the interacting model. \label{Fig5}}
\end{figure}
We now analyze the effect of strong correlations on the above described gap closing and the visibility of quantum oscillations.
Figure \ref{Fig5} shows the comparison of the energy level structure between the interacting and the noninteracting model in an applied magnetic field. The particle number of the $f$ electrons is  fixed to $n_f=1.5$ corresponding to a valence fluctuating regime. We note that in this parameter regime, renormalization effects on the band structure are small even for an interaction strength $U/t=7$, which corresponds to $U/t_f=35$.  

The energy level structures of the  models are obtained from the maximum position of the peaks in the local Green's function in the bulk. Clearly visible in Fig. \ref{Fig5} is the gap at $B/t=0$ around the Fermi energy $E/t=0$ and the Landau levels periodically approaching the Fermi energy. 
By increasing the magnetic field strength, Landau levels from above and below the Fermi energy approach $E/t=0$ and finally close the gap at $B_c/t\approx 4.5$. 
The mechanism of the gap closing is the same as described for the noninteracting continuum model.

Thus, Landau levels cross the Fermi energy for very strong magnetic fields. However, in SmB$_6$ and YbB$_{12}$ quantum oscillations are observed even in the insulating regime, i.e. before the magnetic breakdown. 
\begin{figure}[t]
\begin{center}
    \includegraphics[width=0.9\linewidth]{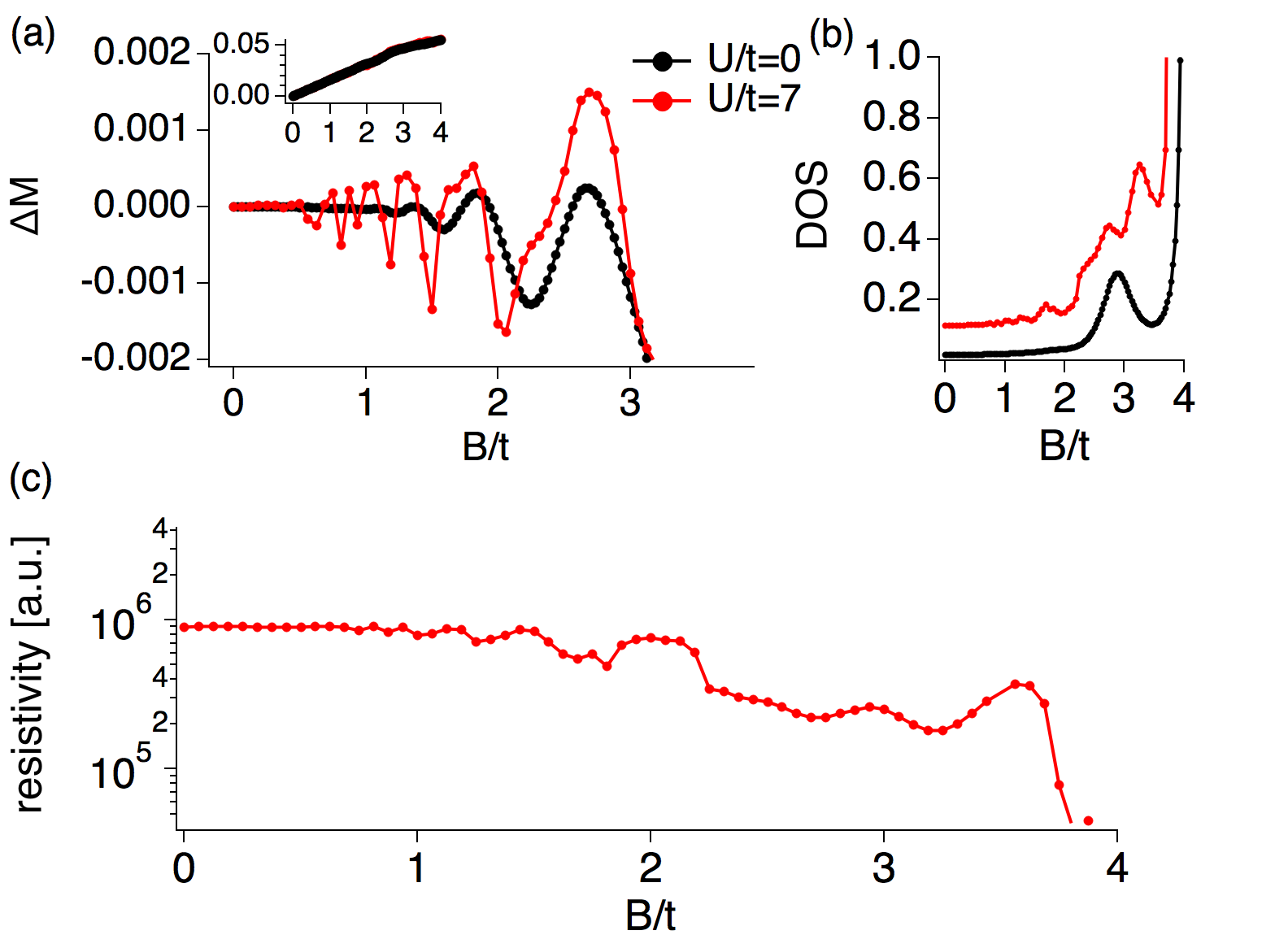}  
\end{center}
\caption{Quantum oscillations in the valence fluctuating regime, $n_f=1.5$. (a) Magnetization (inset) and difference from the linear behavior $\Delta M=M-\alpha B$ of the $f$ electrons; (b) DOS at the Fermi energy; (c) Resistivity for the interacting system.\label{Fig6}}
\end{figure}
We therefore show (a) the magnetization, (b) the density of states (DOS),  and (c) the resistivity in Fig. \ref{Fig6}. The magnetization of the $f$ electrons increases approximately linearly  until the magnetic breakdown. A precise analysis of the magnetization by subtracting the linear part, however, reveals that it includes oscillations shown in Fig. \ref{Fig6}(a). These oscillations occur well below the magnetic breakdown. Much more importantly, these oscillations are strongly enhanced by the correlations. While for the noninteracting model, the oscillations can be observed for $B/t>1.5$, in the interacting model they are visible for $B/t>0.5$. Directly in the DOS, which is small compared to the DOS at the magnetic breakdown, we can also see several small oscillations in the interacting model, see Fig. \ref{Fig6}(b). Finally, we show the resistivity in Fig. \ref{Fig6}(c) calculated by the  Kubo formula within the DMFT approximation, which neglects vertex corrections. To use the vector potential $A=B(-y,0,0)$, we have introduced different layers in the $y$-direction in our real-space DMFT calculation. In the $x$-direction, the lattice still has periodic boundary conditions. We  can thus calculate the conductivity in the $x$-direction using the Kubo-formula in the same way as done for superlattices.\cite{PhysRevB.88.155134}

We only show the results for the interacting system, because the resistivity of the noninteracting system strongly depends on an artificial broadening which must be included to obtain a finite resistivity. In the interacting system, such an artificial broadening is unnecessary  because of the self-energy calculated by the DMFT. We see that the resistivity is large, as can be expected for an insulating material. Because of the gap closing, the resistivity decreases with increasing magnetic field. Furthermore, oscillations are clearly visible. We note that the numerical error can be expected to be constant for all magnetic field strengths. Thus, the oscillations observed at high magnetic fields cannot be attributed to numerical errors, as demonstrated by the smooth behavior of the observables at low magnetic fields. We thus conclude that even in these calculations which do not exhibit a strong renormalization effect because of the particle number, $n_f=1.5$, we nevertheless can observe oscillations in the magnetization and the resistivity which are clearly enhanced, compared to the noninteracting system. We note that the resistivity drops by two orders of magnitude to values below $10^3$ at the magnetic breakdown, $U/t=4.5$. This confirms that the quantum oscillations occur in the insulating regime.

\begin{figure}[t]
\begin{center}
    \includegraphics[width=0.9\linewidth]{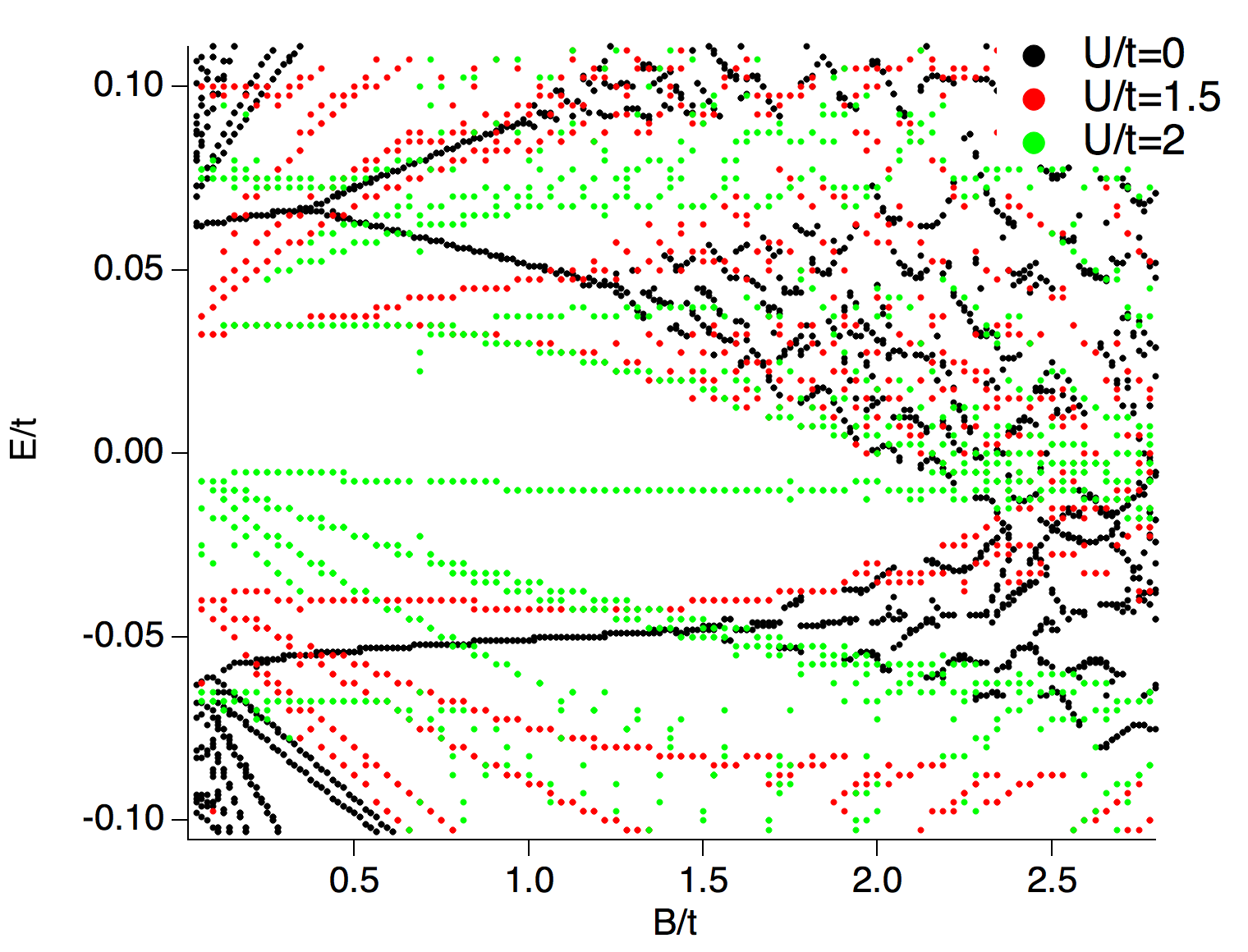}  
\end{center}
\caption{Energy diagram in the Kondo regime, $n_f=1.04$. Comparison of the energy level structure between noninteracting and interacting model.  \label{Fig7}}
\end{figure}

\begin{figure}[t]
\begin{center}
    \includegraphics[width=0.9\linewidth]{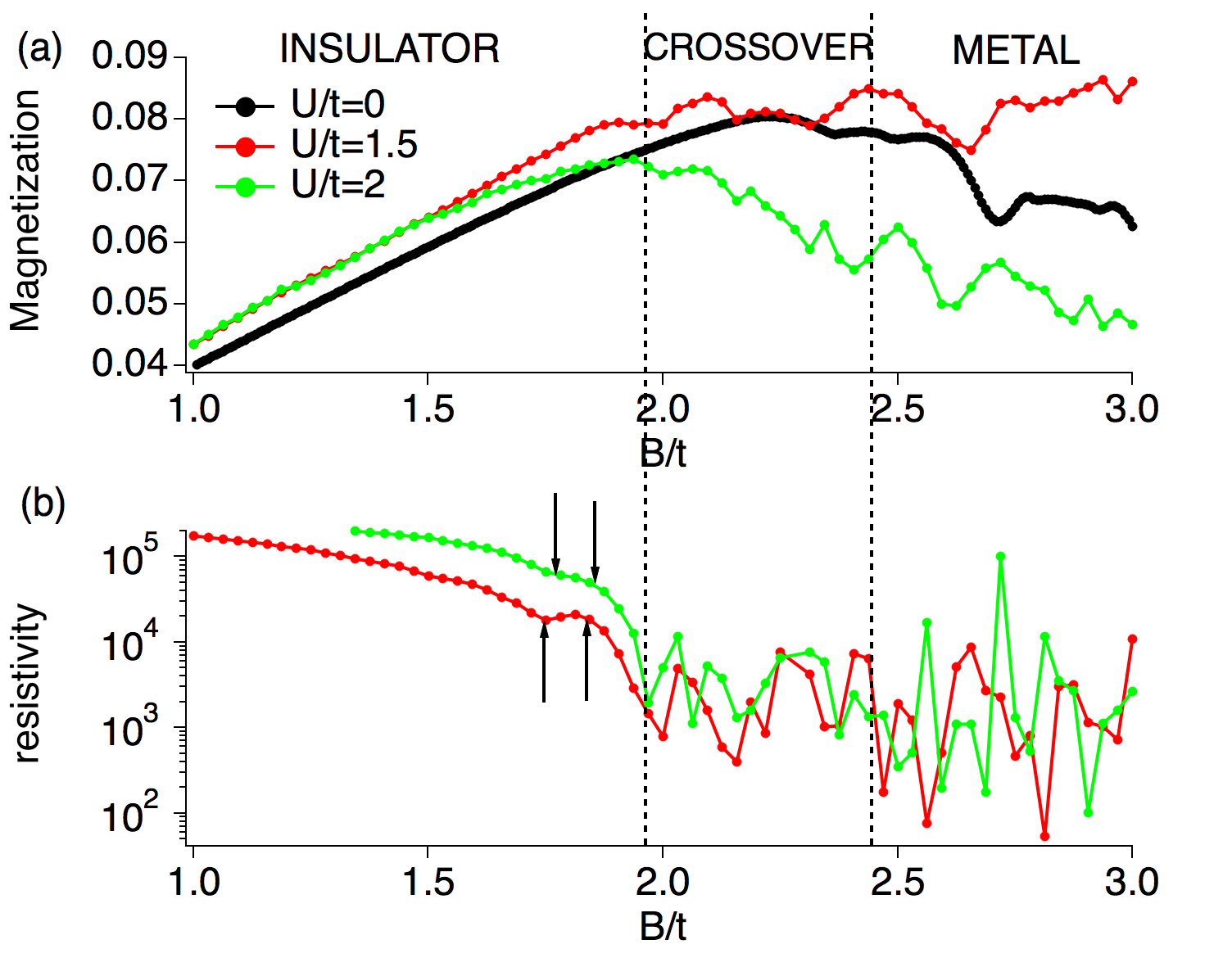}  
\end{center}
\caption{Quantum oscillations in the Kondo regime, $n_f=1.04$. The panels show the magnetization of the $f$ electrons in (a),  and the resistivity in (b) for different interaction strengths. We note that the resistivity of $U/t=2$ is shifted by an positive offset for reasons of clarity. Arrows denote small oscillations in the regime with  high resistivity.
\label{Fig8}}
\end{figure}

Although the calculations in Figs. \ref{Fig5} and \ref{Fig6} are performed for strong interaction, there is only a weak renormalization because correlation effects are suppressed away from $n_f=1$ in our model. To better understand the effect of strong interactions and renormalization, we show in Figs. \ref{Fig7} and \ref{Fig8} calculations in the Kondo regime with particle number $n_f=1.04$. 
For this particle number, the interaction strength $U/t=1.5$ already leads to a considerable renormalization as shown in Fig. \ref{Fig7}. Increasing the interaction strength to $U/t=2$, we observe that the gap at zero magnetic field is smaller than $1/3$ of the noninteracting band gap. Irrespective of the interaction strength, we see that Landau levels approach the Fermi energy and close the gap at a critical magnetic field strength, $B_c/t\approx 2.4$.
The renormalization has thereby two important effects. First, the gap at zero field becomes smaller. Second, the slope of the gap closing, which strongly depends on the mass of the particles involved, also becomes  smaller. Thus, the critical field strength of the gap closing is essentially unchanged from the noninteracting case. 

Figure \ref{Fig8} shows the magnetization and the resistivity in the Kondo regime over a wide range of magnetic fields including the insulating and the metallic regimes. As can be confirmed in Fig. \ref{Fig7}, Landau levels cross the Fermi energy for magnetic field strengths larger than $B/t\approx 2.4$. Thus, the system shows normal metallic behavior for these magnetic field strengths. However, already for magnetic field strengths $2<B/t< 2.4$, the gap becomes very small.
This is confirmed by the resistivity which strongly decreases around $B/t\approx 2$. However, we note that the values of the resistivity for $B/t>2.4$ when a Landau level crosses the Fermi energy are another order of magnitude lower than the resistivity for magnetic fields $2<B/t< 2.4$. We thus define three regions in our results: An insulating region with high resistivity, $B/t<2$, a crossover region with intermediate resistivity, $2<B/t<2.4$, and a metallic region $B/t>2.4$.
Already in the insulating region, we observe small oscillatory behavior in the calculated properties. We note that by an analysis as done in Fig. \ref{Fig6} the number of visible oscillations increases.
Landau levels are still energetically well separated from the Fermi energy in this regime and thus do not have big influence on the physical properties of the system. In the crossover region, when Landau levels come very close to the Fermi energy but do not yet cross it,  Landau levels have a strong impact on the physical properties. We observe quantum oscillations with large amplitude in the resistivity and the magnetization of the system. The amplitude of these oscillations is much larger than in the noninteracting system. As before, we do not show the resistivity for the noninteracting system, because it becomes only finite when inserting an artificial and arbitrary life-time of the particles. In the metallic region for magnetic  fields larger than the critical field strength, the amplitude of the quantum oscillations further increases.

It is remarkable that quantum oscillations with large amplitude can be observed in the crossover region although Landau levels do not reach the Fermi energy. The reason for this phenomenon is the finite-life time of the quasiparticles in the Landau levels induced by the correlation effects (imaginary part of the self-energy). We furthermore note that in the experiments for SmB$_6$ and YbB$_{12}$ the resistivity decreases before quantum oscillations are observed. This could be evidence for the narrowing of the gap with increasing the magnetic field strength as also found in our analysis.

\begin{figure}[t]
\begin{center}
      \includegraphics[width=\linewidth]{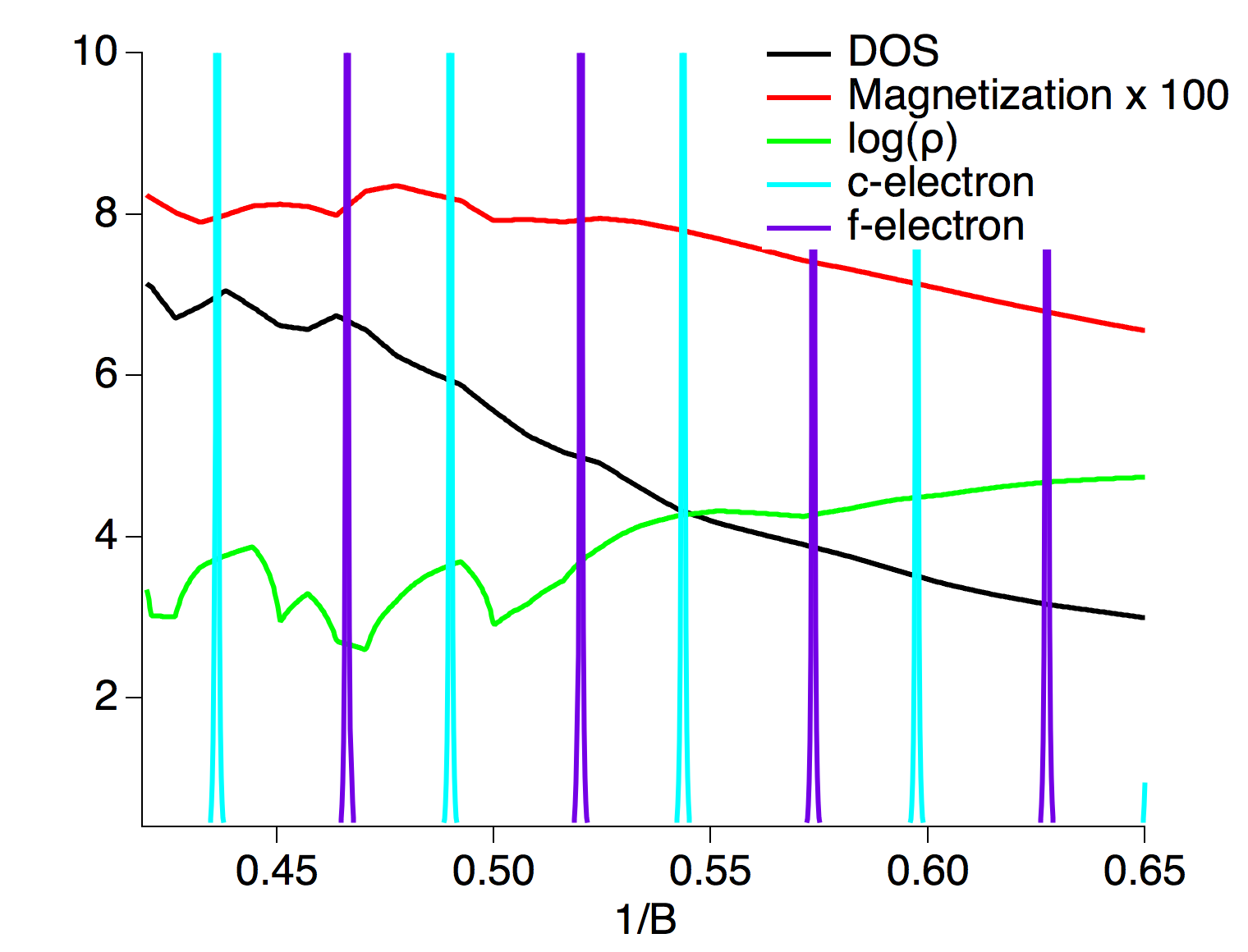}  
    \end{center}
\caption{Quantum oscillations in the DOS, the magnetization, and the resistivity in the Kondo regime ($n_f=1.04$) shown for $1/B$, $U/t=1.5$. The figure shows the crossover regime, $1/B<0.5$, and the insulating regime, $1/B>0.5$.
We furthermore include the peaks of the $c$ and $f$ electron Landau levels simulated by equation (\ref{EQ:LandauLevel}). The magnetic breakdown occurs at $1/B=0.42$.
\label{Fig9}}
\end{figure}
Although these oscillations occur before the gap closes, Landau levels of the $c$ and $f$ electrons approaching the Fermi energy are responsible for these quantum oscillations.
As can be seen in Figs. \ref{Fig5} and \ref{Fig7},  Landau levels come very close to the Fermi energy. 
In Fig. \ref{Fig9}, we show again the quantum oscillations in the DOS, the magnetization, and the resistivity in the Kondo regime, $n_f=1.04$, for $U/t=1.5$ plotted as $1/B$. Furthermore, we include into this figure the delta-peaks of the $c$ and $f$ electron Landau levels. These peaks correspond to the magnetic field strengths where the unhybridized $c$ and $f$ electron Landau levels, described by Eq. (\ref{EQ:LandauLevel}), cross the Fermi energy. Although there is no perfect agreement between the quantum oscillations in the observed quantities and the peaks of the Landau levels, we see that the frequency of these Landau levels can explain the frequencies seen in the quantum oscillations of the DOS, the magnetization, and the resistivity. 
This proves that we can understand the quantum oscillations in strong magnetic fields before the magnetic breakdown as correlated Landau level approaching the Fermi energy. Correlations, which results in renormalization and broadening of the Landau levels, thereby strongly enhance the amplitude of the quantum oscillations in the observed quantities.

\section{three-dimensional system}\label{threeD}
\begin{figure}[t]
\begin{center}
      \includegraphics[width=\linewidth]{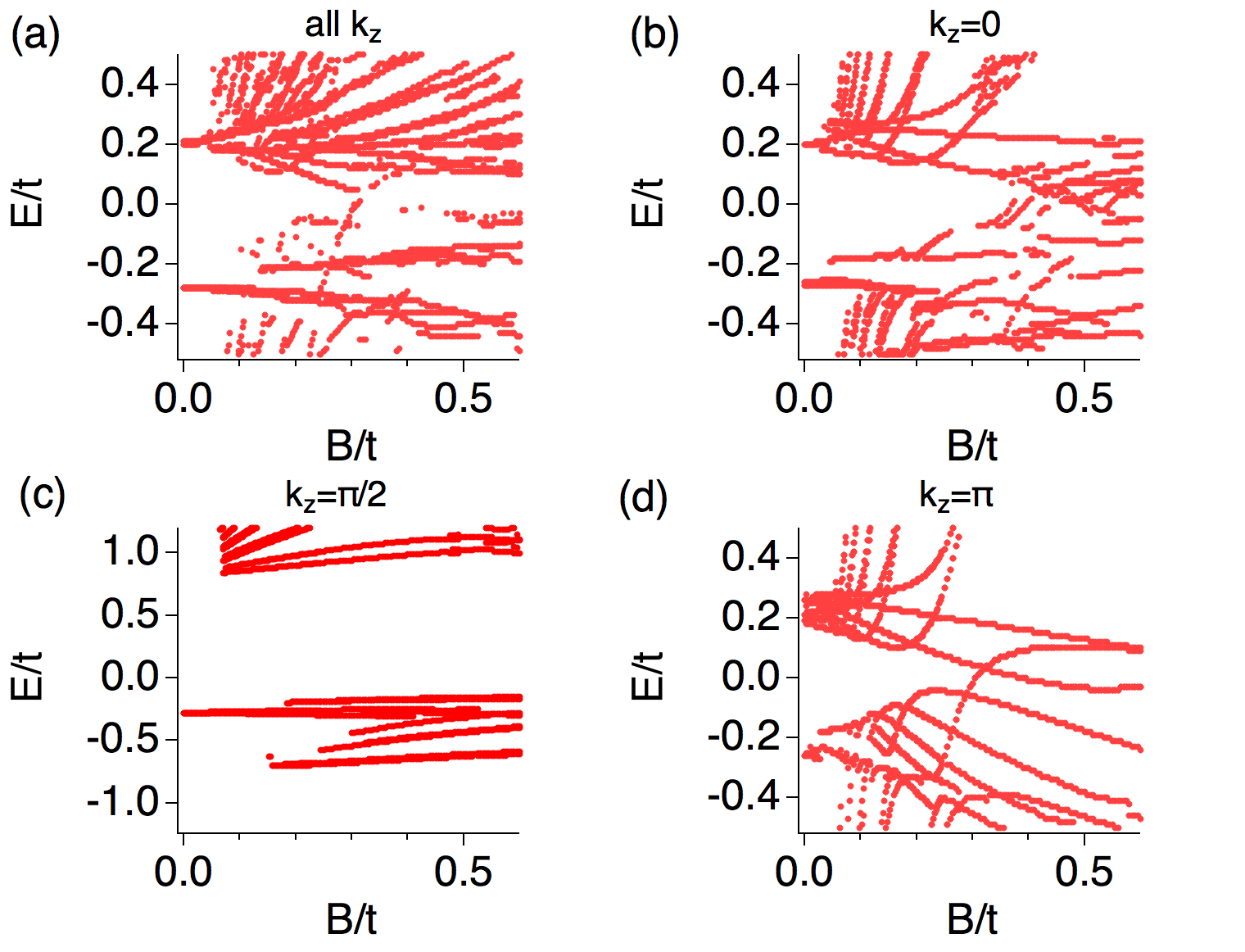}  
    \end{center}
    \caption{Landau level structure for the 3D model for $U/t=6$, $V/t=0.2$ in the valence-fluctuating regime. Analysis of the Landau level structure of the converged 3D solution for all $k_z$ (a), $k_z=0$ (b), $k_z=\pi/2$ (c), and $k_z=\pi$ (d). \label{Fig10}}
\end{figure}
\begin{figure}[t]
\begin{center}
      \includegraphics[width=\linewidth]{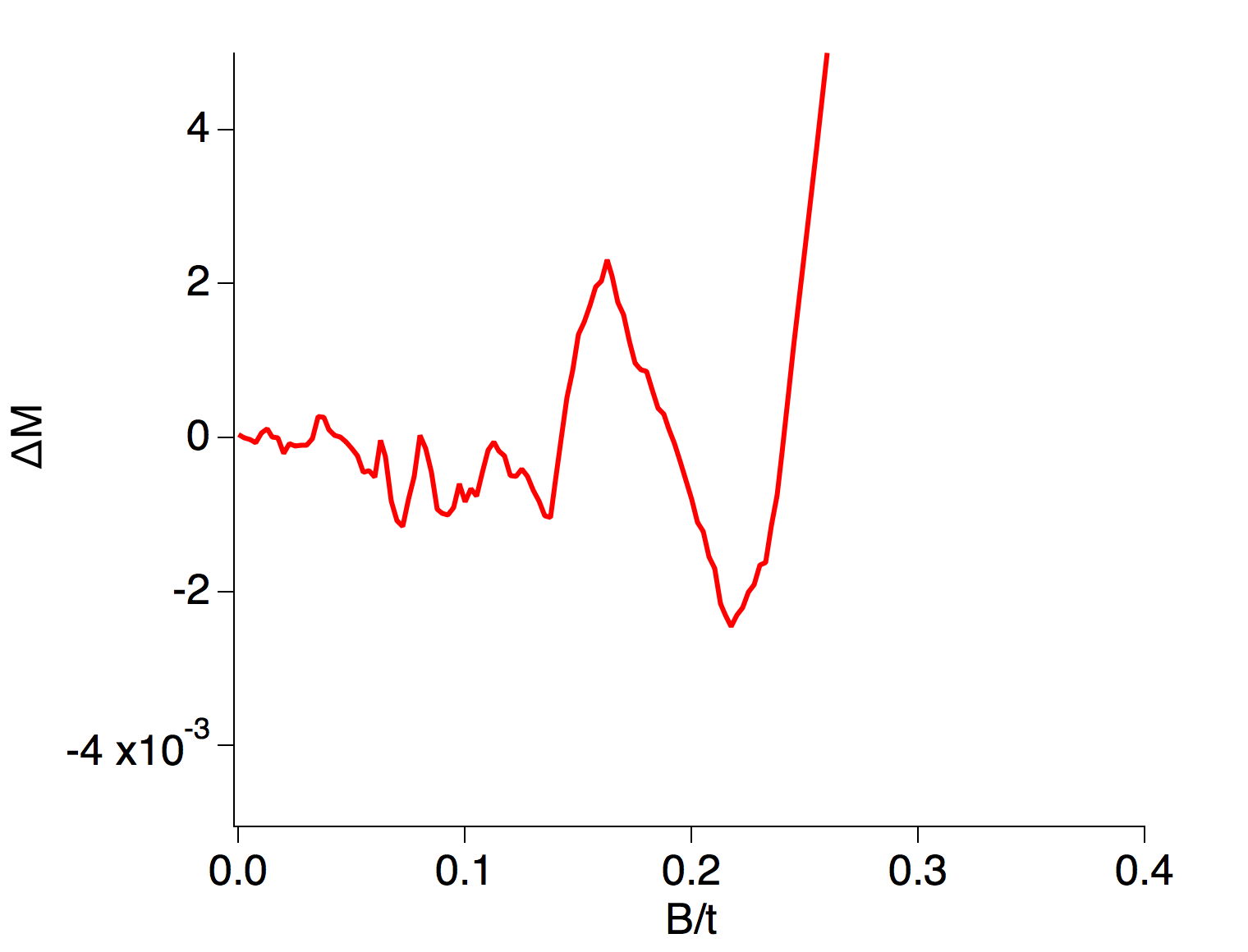}  
    \end{center}
   \caption{ Magnetization of the $f$ electrons in the 3D model for $U/t=6$ in the valence-fluctuating regime. We only show the results after subtracting a quadratic fit to enhance the visibility of the oscillatory behavior. The magnetic breakdown occurs around $B/t=0.3$. \label{Fig11}}
\end{figure}

Up to now, we have analyzed a 2D model of a topological Kondo insulator, where the magnetic field is perpendicular to the lattice. However, SmB$_6$ and YbB$_{12}$ are 3D materials. Electron hopping and hybridization in the direction of the magnetic field are not influenced by the Landau quantization. Thus, the hybridization in the direction of the magnetic field can open a gap between the $c$- and $f$-electron bands. As a consequence, the gap-closing mechanism due to the nonlocal hybridization and the quantum oscillations due to Landau levels approaching the Fermi energy as described above are influenced and might even not occur at all
because of the hybridization in the direction of the magnetic field. 

However, we here show that the above-mentioned physics remains essentially intact and the 3D system can be understood in the same way as the 2D model.
For this purpose we use a 3D model of a topological Kondo insulator.\cite{PhysRevB.98.075104}
The Hamiltonian reads
\begin{eqnarray*}
H&=&H_0+H_{\mathrm{int}},\\
H_0&=&\sum_k\sum_{\sigma=\{\uparrow,\downarrow\}}\sum_{o=\{c,f\}}\epsilon^o_kc^\dagger_{k,\sigma,o}c_{k,\sigma,o}\\
&&+V\sum_{k,\tau_1,m\tau_2}c^\dagger_{k,\tau_1,c}c_{k,\tau_2,f}\sin k_x\sigma^x_{\tau_1\tau_2}\\
&&+V\sum_{k,\tau_1,m\tau_2}c^\dagger_{k,\tau_1,c}c_{k,\tau_2,f}\sin k_y\sigma^y_{\tau_1\tau_2}\\
&&+V\sum_{k,\tau_1,m\tau_2}c^\dagger_{k,\tau_1,c}c_{k,\tau_2,f}\sin k_z\sigma^z_{\tau_1\tau_2}\\
&&+0.2\sum_{i,\sigma}n_{i,\sigma,c},\\
\epsilon^c_k&=&-0.1(\cos(k_x)+\cos(k_y)+\cos(k_z))\\
&&+0.075\cos(k_x)\cos(k_y)\\&&+0.075\cos(k_y)\cos(k_z)\\&&+0.075\cos(k_x)\cos(k_z)\\
&&+0.15\cos(k_x)\cos(k_y)\cos(k_z),\\
\epsilon^f_k&=&-0.1\epsilon^c_k,\\
H_{\mathrm{int}}&=&U\sum_i n_{i,\uparrow,f}n_{i,\downarrow,f}.
\end{eqnarray*}
The operator $c^\dagger_{k,\sigma,o}$ creates an electron with momentum $k$, spin direction $\sigma$ in orbital $o\in\{c,f\}$. $\epsilon^o_k$ describes the energy depending on the momentum for each orbital. The energies have been chosen in a way that there are band inversions between $c$-electrons and $f$-electrons at $(\pi,0,0)$, $(0,\pi,0)$, and $(0,0,\pi)$ in the Brillouin zone, which resembles qualitatively the band structure of SmB$_6$. We include nearest neighbor, next-nearest neighbor and next-next-nearest neighbor hopping on a cubic lattice.
Due to the hybridization, $V$, between the $c$-electron band and the $f$-electron band, a gap opens in the bulk spectrum.  $\sigma^x$, $\sigma^y$, $\sigma^z$ are the Pauli matrices.
The operator  $n_{i,\sigma,c}$ and $n_{i,\sigma,f}$ are local density operators on lattice site $i$ for the $c$-electrons and $f$-electrons, respectively. Finally, $H_{\mathrm{int}}$ describes a repulsive local density-density interaction in the $f$-electron band,  which is necessary to describe the Kondo effect in strongly interacting $f$-electron systems.  
We have used the same model to analyze the interplay between magnetism and topology in a topological Kondo insulator in our previous work.\cite{PhysRevB.98.075104}. We again solve this model by means of real-space DMFT. Compared to the 2D model, the additional dimension shows up when calculating the local Green's functions which stipulate the input of the impurity models. We use again the  vector potential $\vec A=B(-y,0,0)$ for our calculations.

In Figs. \ref{Fig10} and \ref{Fig11}, we show results for the valence-fluctuating regime, $n_f=1.6$, for  $U/t=6$ and $V/t=0.2$. In Fig. \ref{Fig10}, we again extract the Landau level position by analyzing the peaks in the Green's function of the bulk system. Figure \ref{Fig10}(a) shows the Landau level structure for the local bulk Green's function, where we have integrated over $k_z$. We clearly see that Landau levels approach the Fermi energy and close the gap at a critical magnetic field strength, $B/t\approx 0.3$. As explained above, this might be surprising as the hybridization in $z$-direction acts between Landau levels with the same index and is not influenced by the magnetic field at all.  

To understand the Landau level structure in Fig. \ref{Fig10}(a), we show the level structure for separate momenta $k_z$ in (b)-(d). Here it becomes clear that depending on $k_z$ a gap at the Fermi energy exists or not. This can be understood by the fact that in a topological insulator the hybridization is momentum dependent and has nodes for which the hybridization vanishes. In our 3D model, the hybridization in $z$-direction is proportional to $\sin k_z$ and thus vanishes for $k_z=0$ and $k_z=\pm\pi$.  For these momenta, only Landau levels with different index hybridize and the above-described physics for the 2D system holds true. It can be clearly seen that for $k_z=0$ [Fig. \ref{Fig10}(b)] and $k_z=\pm\pi$ [Fig. \ref{Fig10}(d)] Landau levels approach the Fermi energy and close the gap in the same way as in the valence-fluctuating regime in the 2D model. On the other hand, for $k_z=\pi/2$ [Fig. \ref{Fig10}(c)] the hybridization in $z$-direction does not vanish and we observe a gap at the Fermi energy. Finally, the local Green's function as shown in Fig. \ref{Fig10}(a) is the integral/sum over all $k_z$. Thus, the physics at the Fermi energy is determined by the momenta where the hybridization in the direction of the magnetic field vanishes, because for these momenta Landau levels approach the Fermi energy and the gap is closed in the same way as in the 2D system. 
We note that the resolution of the figure is lower than for the 2D system, as the calculations are numerically more demanding. This leads to a low resolution of the structure: not all the points found in the level structure in Figs. \ref{Fig10} (b)-(d) can be seen in Fig. \ref{Fig10}(a).

Having established the existence of the gap closing in strong magnetic fields, which is similar to the 2D model, it is not surprising to observe quantum oscillations in the magnetization before the gap closes, which is shown in Fig. \ref{Fig11}. To enhance the clarity of the figure, we have subtracted a quadratic background. We clearly observe oscillatory behavior before the gap closes at $B/t=0.3$.

\section{discussion and conclusions\label{sec_dis}}
\begin{figure}[t]
\begin{center}
      \includegraphics[width=\linewidth]{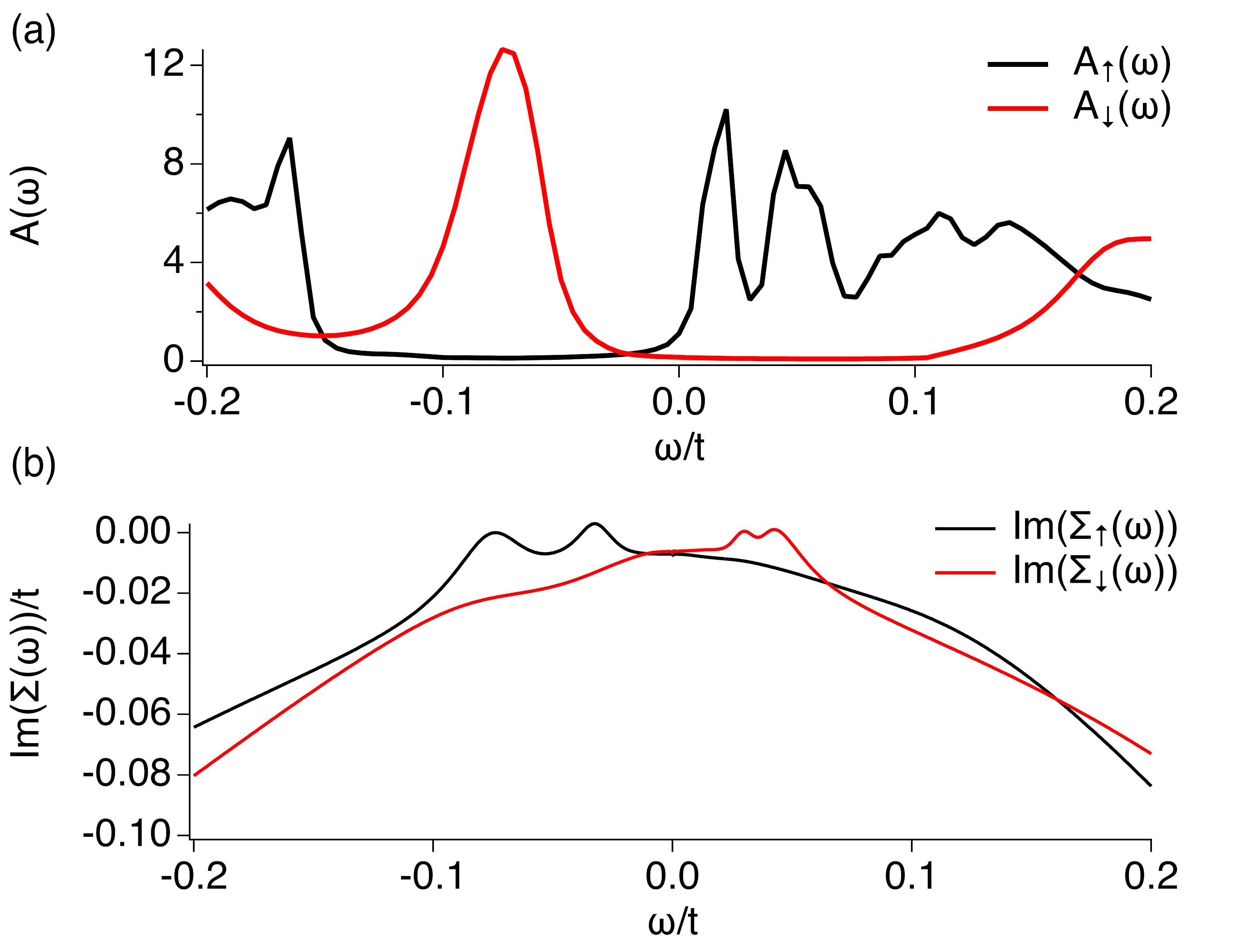}
    \end{center}
\caption{Local DOS (a)  and imaginary part of the self-energy (b) for $B/t=2$, $U/t=1.5$ in the Kondo regime. The DOS clearly includes separated Landau levels very close to the Fermi energy, $\omega/t=0$. The self-energy becomes small but is finite close to the Fermi energy which results in a broadening of the Landau levels.}
\label{Fig12}
\end{figure}

Let us discuss these numerical results in more detail. Our calculations have shown that quantum oscillations can be observed in the DOS, the magnetization, and the resistivity of the bulk of a topological Kondo insulator for magnetic fields smaller than the magnetic breakdown.  We note that topological surface states do not contribute to the oscillations shown here. However, we believe that metallic surface states will contribute to the experimental observed quantum oscillations.
Thus, a computational analysis of how topological surface states contribute to the observed quantum oscillations is an interesting question, which is left as a future project.

The  frequencies of the bulk quantum oscillations observed here, when plotting these  quantities over $1/B$, agree well with the frequencies generated by unhybridized  $c$- and $f$-electrons. Thus, the oscillations can be reproduced by taking into account the Landau levels of the light $c$ electrons and those of the heavy $f$ electrons. The Fermi surface which causes these oscillations is small, because the electrons must be treated as unhybridized. However, although the Fermi surface of the $f$ electrons is small, it is important to realize that the $f$ electrons are renormalized and heavy; the mass $\frac{1}{2\pi}\frac{\partial A_f(E)}{\partial E}=m_f^*$ is much larger than that of the $c$ electrons.

These results naturally lead us to propose the notion of a virtual Fermi surface, which is created by the unhybridized $c$ and $f$ electrons. 
This virtual Fermi surface can be observed at high magnetic fields. Because of a hybridization between $c$ and $f$ electrons,  which is an odd function of the momentum  such as $\sigma_i\sin(k_i)$ and thus leads to a coupling of Landau levels with different index, the effect of the hybridization becomes invisible when the cyclotron frequency is of the order of the hybridization strength. The closing of the gap in strong magnetic fields and the observation of quantum oscillations  thus provide evidence for a hybridization with odd momentum dependence, which ubiquitously appears in topological insulators. To observe the gap-closing and the related quantum oscillations in an insulator, this kind of hybridization is a necessary ingredient. 
For 3D materials, the gap closes exactly for the momenta for which the hybridization in direction of the magnetic field between $c$ and $f$ electrons vanishes. The physics at the Fermi energy can be understood by these 2D momentum planes.

Furthermore, we have demonstrated by direct calculation that strong correlations enhance the amplitude and thus the visibility of the oscillations in observable quantities such as the magnetization and the resistivity for magnetic fields smaller than the magnetic breakdown. There are two reasons for this. First, because of the renormalization, the slope of the gap closing and the gap width  are reduced. If we assume that experiments can detect  quantum oscillations of Landau levels which are slightly away (but not too far) from the Fermi energy, then the renormalization will enlarge the range of magnetic fields for which Landau levels are observable. This effect can be seen in Fig. \ref{Fig7} comparing between the noninteracting  and the interacting energy level structure.
Second, because of the self-energy arising from the strong correlations, quasiparticle bands away from the Fermi energy are broadened, and can thus influence observable quantities at the Fermi energy. To demonstrate this effect, we show in Fig. \ref{Fig12}, the DOS and the imaginary part of the self-energy for $U/t=1.5$ and $B/t=2$ in the Kondo regime, $n_f=1.04$.  In the DOS, we see several energetically separated peaks, which correspond to the Landau levels close to the Fermi energy. These Landau levels are broadened and thus the DOS at the Fermi energy is increased, making it possible to observe quantum oscillations at the Fermi energy. The broadening of the Landau levels is thereby given by the imaginary part of the self-energy shown in Fig. \ref{Fig12}(b). This self-energy is self-consistently calculated by DMFT and describes the correlations (which leads to a finite life-time) of the $f$ electrons. We see that although the imaginary part of the self-energy becomes smaller when approaching the Fermi energy, it is still finite and results in the broadening of the Landau levels. 
We note that under certain conditions, the broadening and the renormalizing of
quasiparticles bands due to the self-energy can turn even a band insulator into a metal
\cite{PhysRevLett.97.046403,Hoang_2010,PhysRevB.80.155116,PhysRevB.76.085112}. In this work, however, the self-energy of the $f$ electrons broadens quasiparticle
bands close to the Fermi energy, resulting in a slight increase of the density of states at the
Fermi energy for small magnetic fields. When quasiparticle bands eventually cross the
Fermi energy at large magnetic fields, the DOS increases by several orders of magnitude.
Thus, the resistivity is high for small magnetic fields consistent with an insulator.
Furthermore, we note that if the imaginary part becomes too large (the life-time becomes too short), Landau levels in the DOS merge to broad bands and quantum oscillations would not be observable anymore.

We believe that this scenario can explain the observations in SmB$_6$, which is a good candidate for a topological Kondo insulator, thus including strong correlations and an odd momentum-dependent hybridization. Oscillations are observed for strong magnetic fields but before the magnetic breakdown. Because the material is still insulating in these fields, the oscillations can be understood as strongly correlated Landau levels approaching but not crossing the Fermi energy. Furthermore, the experimentally observed oscillations show characteristics of the metallic rare-earth hexaborids,\cite{Tan287} which do not form a hybridization gap at the Fermi energy. Thus, the conclusion that the oscillations originate in the unhybridized $c$  and $f$ electrons with virtual Fermi surface agrees well with the experimental observations in SmB$_6$.
On the other hand, the case of YbB$_{12}$ seems to be more difficult, as there seems to be no direct correspondence between the observed oscillations in metallic LuB$_{12}$.
However, comparing the band structures of YbB$_{12}$ and LuB$_{12}$ \cite{0953-8984-30-16-16LT01}, it seems possible that the Boron bands are energetically shifted in these compounds, which would explain different quantum oscillations.
Furthermore, because SmB$_6$ and YbB$_{12}$ are topologically nontrivial, it can be expected that metallic surface states contribute to the experimentally observed quantum oscillations. Thus, it remains an interesting problem to distinguish quantum oscillations due to the metallic surface states from those due to the insulating bulk in the experimental data.

In summary, we have shown that quantum oscillations can be observed in topological Kondo insulators for magnetic fields before the magnetic breakdown. 
 While it is difficult to observe quantum oscillations in the noninteracting model, we have  demonstrated that oscillations are strongly enhanced in the correlated model. Thus, strong correlations are essential for the experimental observation of quantum oscillations in an insulator. We  have further shown that the quantum oscillations can be explained by a virtual Fermi surface made of unhybridized light $c$ and heavy $f$ electrons.

\begin{figure}[t]
\begin{center}
      \includegraphics[width=\linewidth]{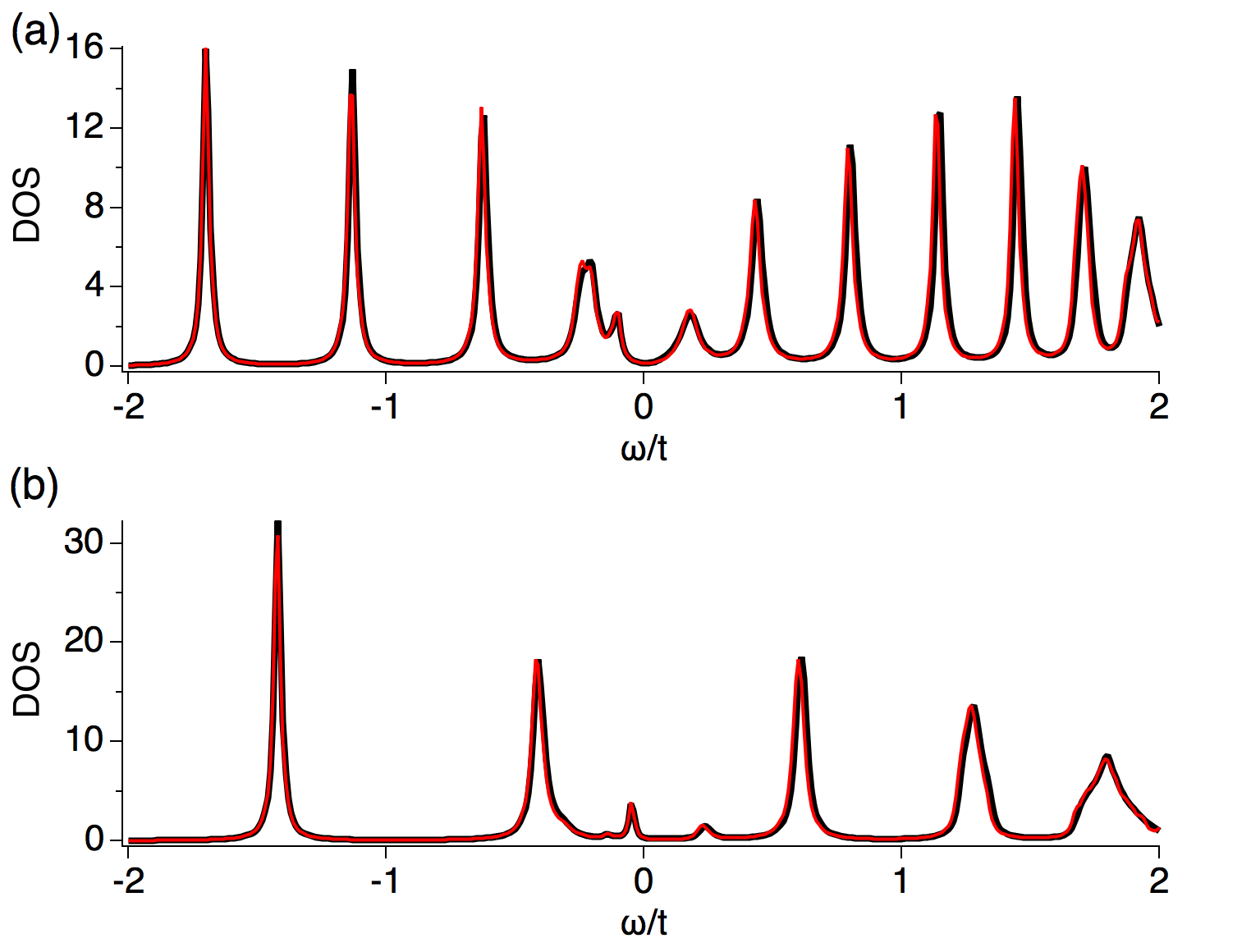} \\
            \includegraphics[width=\linewidth]{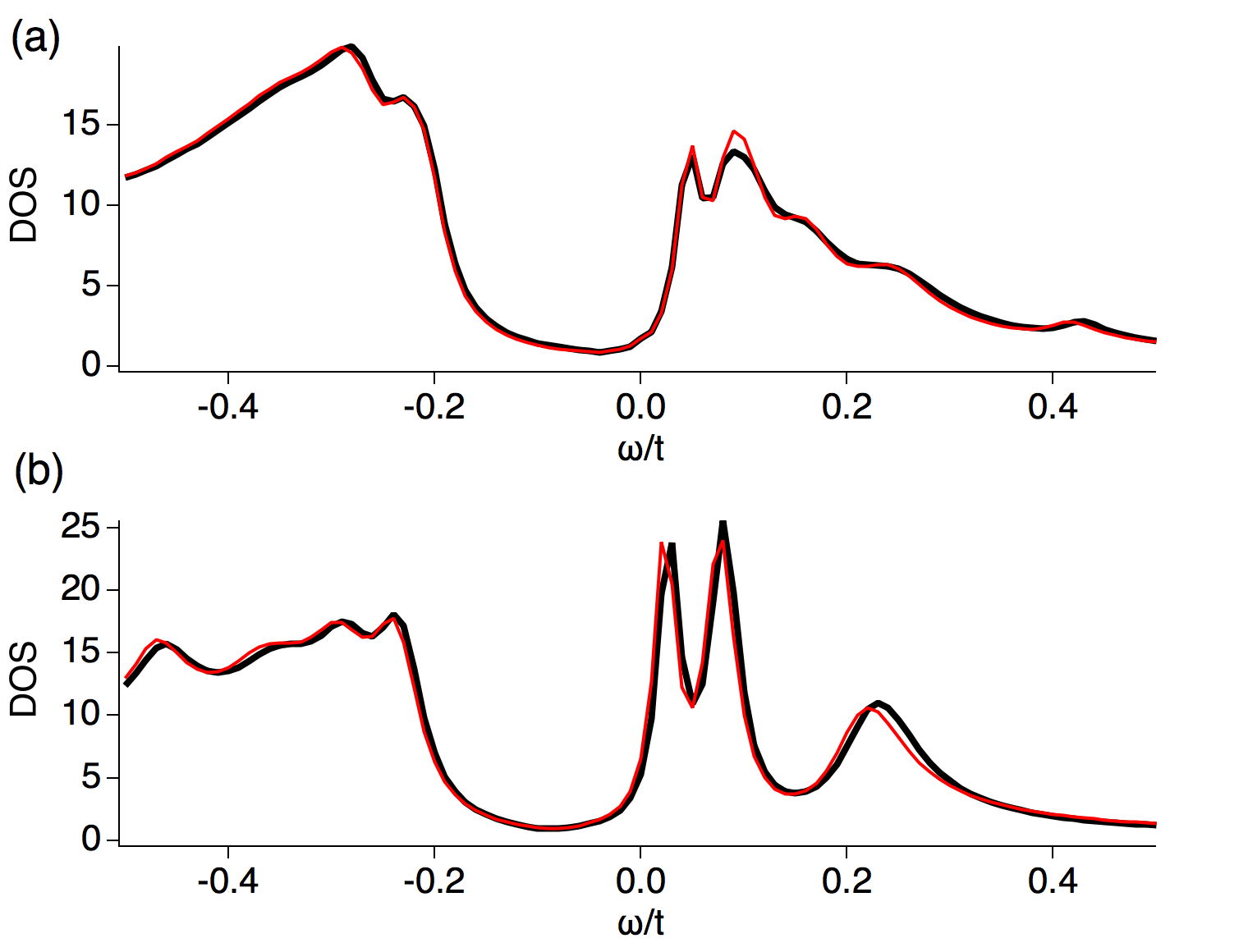} 
    \end{center}
\caption{Local DOS and comparison to the periodic lattice calculation (a) $B/t=\frac{2\pi}{20}/t\sim 1.57$, (b) $\frac{2\pi}{10}/t\sim 3.14$ in the valence fluctuating regime. Black (red) lines corresponds to the periodic boundary (open boundary) calculations. The two panels at the top (bottom) show the $c$ ($f$) electrons.
\label{Fig13}}
\end{figure}

\begin{acknowledgments}
This work is partly supported by JSPS KAKENHI Grant No. 25220711, JP15H05855,  JP16K05501, 18K03511, 18H05842, and No. 18H04316 (JPhysics) and CREST, JST No. JPMJCR1673.
Computer simulations were performed on the "Hokusai" supercomputer in RIKEN and the supercomputer of  the Institute for Solid State Physics (ISSP) in Japan.  \end{acknowledgments}

\appendix
\section{Comparison between periodic and open boundary conditions}
To demonstrate that our approach yields correct results for the magnetic field strengths for which we observe the gap closing, we show in Fig. \ref{Fig13} a comparison between the local density of states in the middle of the slab calculations (open boundary conditions) and the periodic boundary conditions. While the upper panels show the $c$ electron bands for two magnetic field strengths, the lower panels show the $f$ electron band. The local DOS of the $c$ electron band clearly demonstrates the existence of narrow bands, showing up as peaks in the DOS. These narrow bands correspond to the Landau levels. With increasing magnetic field strength the distance between the Landau levels grows. The $f$ electron band, on the other hand, includes rather broad peaks, especially away from the Fermi energy. 
The emergence of broad peaks, instead of very flat bands (narrow peaks), is due to the strong correlations in the $f$ electron band. Strong correlations lead to a finite lifetime of the particles away from the Fermi energy, resulting in broadened bands. Comparing the spectral function calculated using periodic boundary conditions with those calculate with open boundary conditions, we clearly see that there is very good agreement between both calculations. This demonstrates that our calculations with open boundary conditions yield correct results.

\bibliography{paper}


\end{document}